\documentclass[english,prl,twocolumn, aps,amssymb,footinbib,showpacs]{revtex4-1}
\usepackage[T1]{fontenc}
\usepackage{amsmath}
\usepackage{amssymb}
\usepackage{graphicx}
\usepackage{braket}
\usepackage{babel}

\makeatletter

\@ifundefined{textcolor}{}
{
 \definecolor{BLACK}{gray}{0}
 \definecolor{WHITE}{gray}{1}
 \definecolor{RED}{rgb}{1,0,0}
 \definecolor{GREEN}{rgb}{0,1,0}
 \definecolor{BLUE}{rgb}{0,0,1}
 \definecolor{CYAN}{cmyk}{1,0,0,0}
 \definecolor{MAGENTA}{cmyk}{0,1,0,0}
 \definecolor{YELLOW}{cmyk}{0,0,1,0}
 }

\usepackage{bm}\@ifundefined{definecolor}{\usepackage{color}}{}

\usepackage[T1]{fontenc}
\usepackage[latin9]{inputenc}
\usepackage{textcomp}
\usepackage{color}

\usepackage{babel}

\makeatother

\begin{document}

\title{Observing Topological Invariants Using Quantum Walk\\ in Superconducting Circuits}

\author{E. Flurin$^{1,2}$}
\email{emmanuel.flurin@berkeley.edu}
\author{V. V. Ramasesh$^{1,2}$}
\author{S. Hacohen-Gourgy$^{1,2}$}
\author{L. Martin$^{1,2}$}
\author{N. Y. Yao$^{1}$} 
\author{I. Siddiqi$^{1,2}$}
\affiliation{$^{1}$Department of Physics, University of California, Berkeley, CA 94720, U.S.A.}
\affiliation{$^2$Center for Quantum Coherent Science, University of California, Berkeley CA 94720, USA.}

\begin{abstract}
 The direct measurement of topological invariants in both engineered and naturally occurring quantum materials is a key step in classifying quantum phases of matter. Here we motivate a toolbox based on time-dependent quantum walks as a method to digitally simulate single-particle topological band structures.  Using a superconducting qubit dispersively coupled to a microwave cavity, we implement two classes of split-step quantum walks and directly measure the topological invariant (winding number) associated with each. The measurement relies upon interference between two components of a cavity Schr\"odinger cat state and highlights a novel refocusing technique which allows for the direct implementation of a digital version of Bloch oscillations. Our scheme can readily be extended to higher dimensions, whereby quantum walk-based simulations can probe topological phases ranging from the quantum spin Hall effect to the Hopf insulator.
\end{abstract}
\date{\today}
\maketitle

Topological phases elude the Landau-Ginzburg paradigm of symmetry-breaking\ \cite{lifshitz1980statistical}.  Unlike conventional phases, they do not exhibit order parameters that can be locally measured. Rather, their distinguishing features are hidden in quantized, non-local topological invariants, which are robust to all local perturbations\ \cite{moore2010birth,hasan2010colloquium}. While tremendous theoretical progress has been made toward the full classification of topological phases of matter\ \cite{kitaev2009periodic,schnyder2008classification}, a general experimental platform for the direct measurement of topological invariants is lacking.  Here we demonstrate that time-dependent quantum walks comprise a powerful class of unitary protocols capable of digitally simulating single-particle topological bandstructures and  directly observing the associated non-local invariants.

\begin{figure}[h!]
\includegraphics[width=0.44\textwidth]{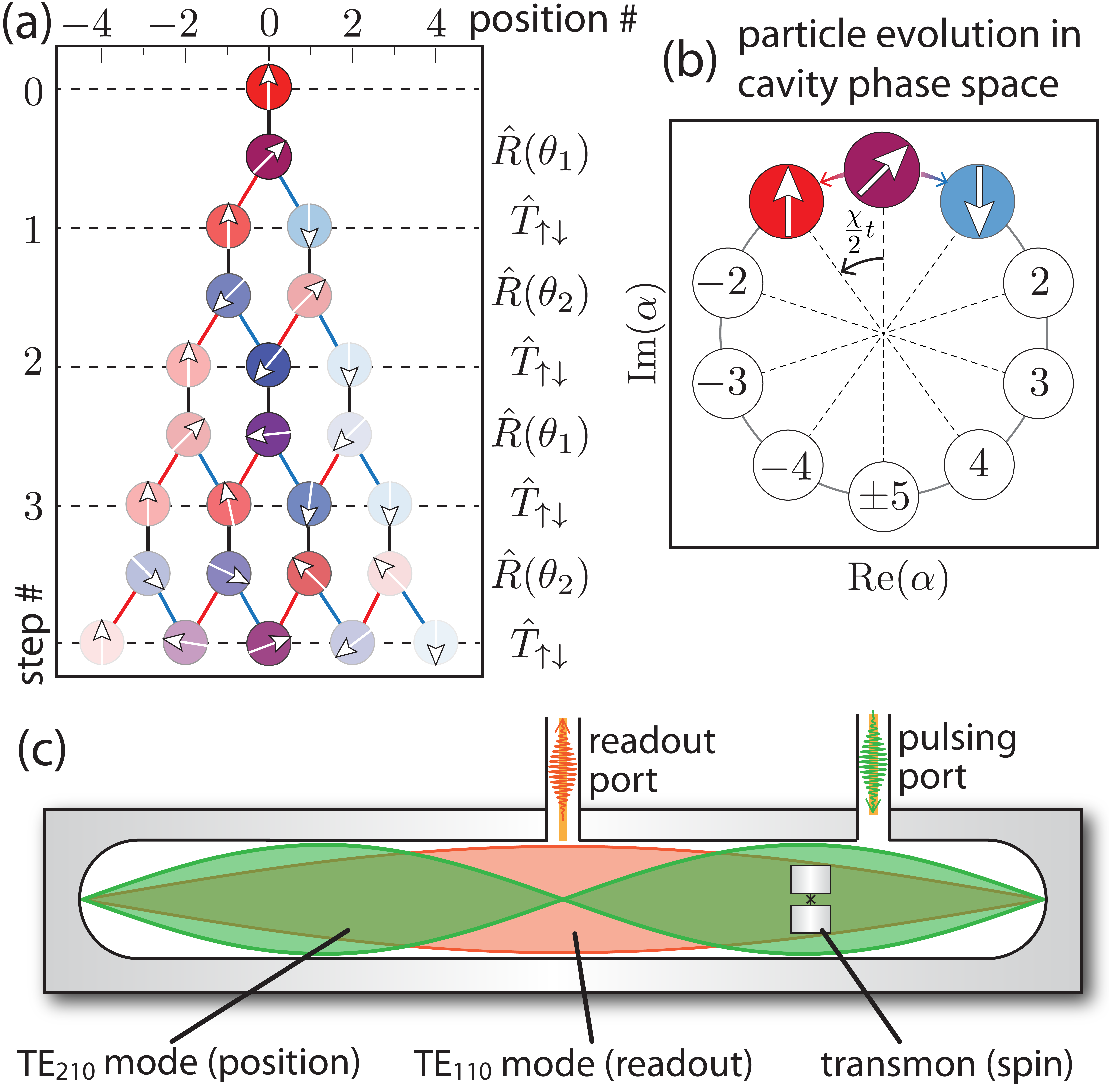}
\caption{\textbf{Quantum walk implementation in cavity phase space.} \textbf{a.} Schematic representation of a split-step quantum walk on a line, with rotations $\hat{R}(\theta_1)$ and $\hat{R}(\theta_2)$ and spin-dependent translation $T_{\uparrow\downarrow}$.  Red (blue) lines show spin-up (-down) components moving left (right).  The opacity of each circle indicates the population on the corresponding lattice site. \textbf{b.} Set of ten cavity coherent states on which the walk takes place, in the phase space of the TE$_{210}$ cavity mode. \textbf{c.} Cavity resonator and qubit.  The fundamental (TE$_{110}$, orange) mode at $\omega_R = 2\pi\times 6.77~\mathrm{GHz}$ is used to measure the qubit state. This mode couples strongly ($\kappa = 2\pi\times600~\mathrm{kHz} = 1/(260~\mathrm{ns})$) to a 50-ohm transmission line via the readout port at the center of the cavity. The TE$_{210}$ cavity mode (green) at $\omega_C = 2\pi\times5.2~\mathrm{GHz}$ is long lived with an inverse lifetime, $\kappa=2\pi\times4~\mathrm{kHz}=1/(40~\mu\mathrm{s})$.  The transmon qubit (coin) has transition frequency $\omega_C = 2\pi\times 5.2~\mathrm{GHz}$, relaxation times $T_1 = 40\ \mu\mathrm{s}$ and $T_2^*=5.2\ \mu\mathrm{s}$, and is dispersively coupled to both cavity modes, with the dispersive shift of the walker mode, $\chi = 2\pi\times 1.61~\mathrm{MHz}$.}
\end{figure}

\begin{figure*}
\includegraphics[width=1\textwidth]{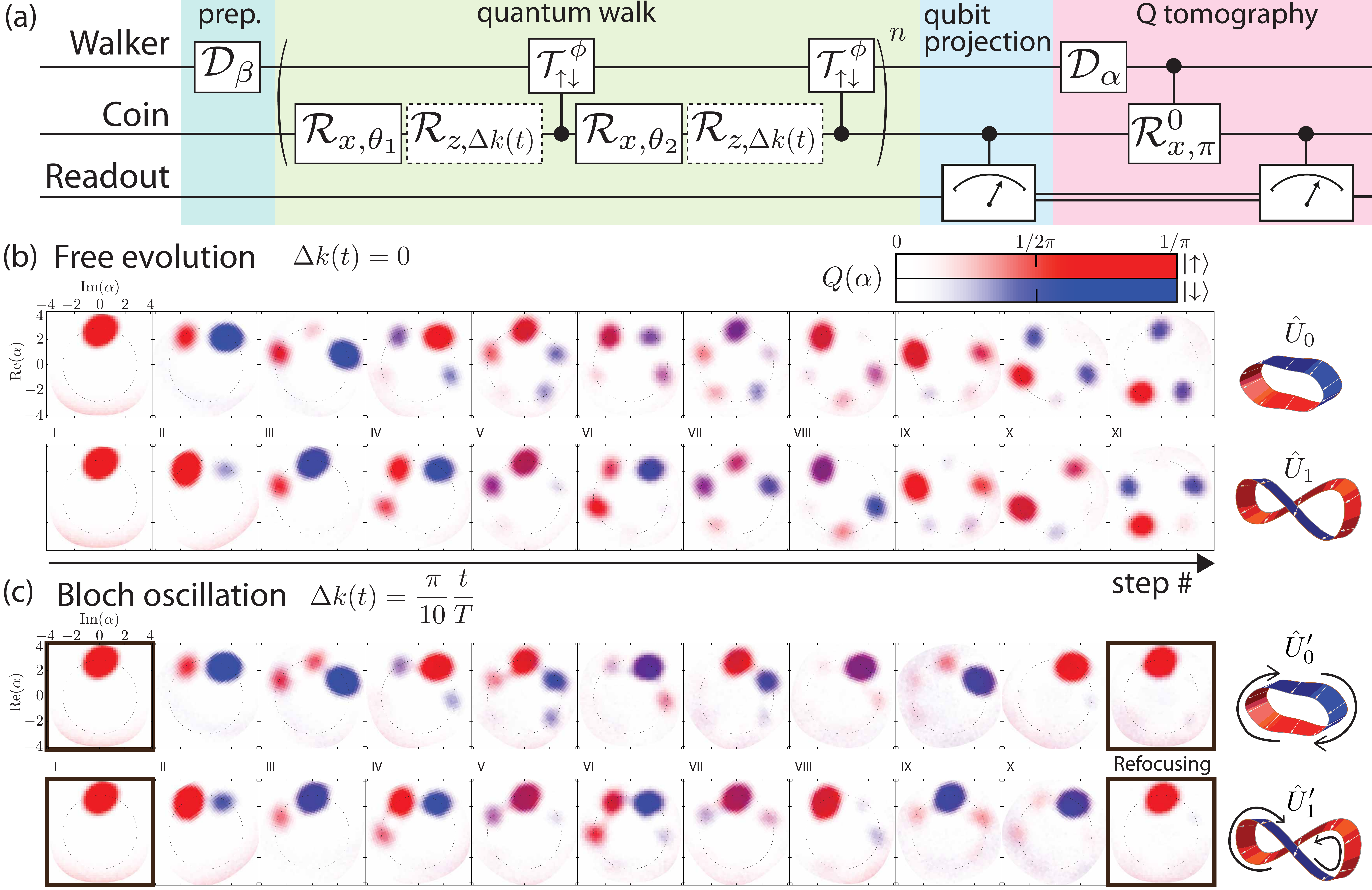}
\caption{\textbf{Quantum walk protocol and resulting populations.} \textbf{a.} Protocol used to perform the quantum walk, showing cavity state preparation (blue), quantum walk (green), qubit state measurement (blue), and Q function measurement (pink).  The dashed boxes with $\sigma_z$ gates are performed to implement the Bloch oscillation. \textbf{b.} Cavity Q functions after each step of the quantum walk without Bloch oscillations, $\hat{U}_0$ (top strip) and $\hat{U}_1$ (bottom strip).  Spin-up (red) and spin-down (blue) Q functions are superimposed.  Average fidelity of the populations compared to theoretical predictions is 0.97 and 0.96 for $\hat{U}_0$ and $\hat{U}_1$, respectively. \textbf{c.} Cavity Q functions after each step of the refocusing quantum walk with Bloch oscillations.  The state refocuses after ten steps, as shown in the final frame for both $\hat{U}_0$ and $\hat{U}_1$.  Refocusing fidelities (to the initial state) for $\hat{U}_0$ and $\hat{U}_1$ are 0.83 and 0.87, respectively.}
\end{figure*}

A quantum walk\ \cite{Y.AharonovL.Davidovich1993,kempe2003quantum,Preiss2015,Farhi1998,Ambainis2004,Childs2009}  describes the  motion of a particle with internal (spin) degrees of freedom moving on a discrete lattice. 
Formally, the quantum walk is comprised of two unitary operations (see Fig.~1A): a \emph{coin toss}, denoted $\hat{R}(\theta)$, which rotates the spin state; and a \emph{spin-dependent translation}, denoted $\hat{T}_{\uparrow\downarrow}$, which translates the particle's position by a single lattice site in a direction determined by the internal spin state. In our cavity quantum electrodynamics implementation of the quantum walk, the particle is encoded as a coherent state of an electromagnetic cavity mode\ \cite{Paik2011} where its position is defined in the cavity's phase space, as shown in figure 1B. Its spin degrees of freedom are formed by a superconducting transmon qubit\ \cite{Koch2007} with basis states $\{\left|\uparrow\right\rangle,\left|\downarrow\right\rangle\}$. To enable the qubit state to control the direction of motion of the coherent state, we realize a strong dispersive coupling between the cavity and  qubit,
\begin{equation}
\hat{H}/\hbar =  \omega_q \hat{\sigma}_z/2 + \omega_c \hat{a}^\dagger\hat{a}-\chi_{qs}\hat{a}^\dagger\hat{a}\hat{\sigma}_{z}/2,
\end{equation}
where $\omega_{q,s}$ are  the qubit and cavity transition frequencies respectively, $\hat{a}$ ($\hat{a}^\dagger$) is the lowering (raising) operator for the cavity mode, $\sigma_z$ the Pauli $z$-matrix for the qubit levels, and $\chi_{qs}$ the dispersive interaction strength (see Fig.~1C).  Dispersive coupling produces a qubit dependent shift in the cavity oscillation frequency.  Viewed in the rotating frame of the cavity at $\omega_c$, the dispersive interaction causes the coherent state to move clockwise (counterclockwise) at a rate $\chi_{qs}/2$ through phase space when the qubit is in the $\left|\uparrow\right\rangle$ ($\left|\downarrow\right\rangle$) state. Thus, free evolution under the dispersive interaction precisely enables the spin-dependent translation needed for the quantum walk\ \cite{xue2008quantum, travaglione2002milburn}.

We realize a particular class of quantum algorithm known as the \emph{split-step} quantum walk\ \cite{Kitagawa2010,kitagawa2012observation}, which alternates two coin tosses (with rotation angles $\theta_1$ and $\theta_2$) between two spin-dependent translations, so that each step of the walk consists of the unitary operation $\hat{U}_W\left(\theta_1,\theta_2\right) =  \hat{T}_{\uparrow\downarrow}\hat{R}(\theta_2)\hat{T}_{\uparrow\downarrow}\hat{R}(\theta_1)$ (see Fig. 1A).  The coin toss operations $\hat{R}_x(\theta)=e^{i\theta \hat{\sigma}_x/2}$ are applied via short ($7.5$~ns) coherent microwave pulses resonant with the qubit transition.  By waiting for a time interval $t = 2\pi(10\chi_{qs})^{-1} =124$~ns between successive coin tosses, we allow the dispersive coupling to naturally implement the spin-dependent translation.  This time interval determines the lattice on which the walk takes place; here, it is a circular lattice of ten sites in cavity phase space  (Fig. 1B).

We begin by performing a pair of topologically distinct split-step quantum walks, the first (topologically trivial) with unitary $\hat{U}_0 = \hat{U}_W\left(3\pi/4,\pi/4 \right)$, and the second (topologically non-trivial) with  $\hat{U}_1 = \hat{U}_W\left(\pi/4,3\pi/4 \right)$. To demonstrate the robustness of the winding number, we also implement an additional pair of walks which are continuously connected to $\hat{U}_0$ and $\hat{U}_1$ (e.g. without closing the gap).
The experimental sequence  is shown in Fig. 2A.  The cavity mode is initialized ($\mathcal{D}_\beta$) in a coherent state $|\beta\rangle$ with $|\beta|^2 = 8$ photons, after which the walk unitary  is repeatedly applied.
To directly reconstruct the walker's quantum state on the phase space lattice, we first projectively measure the qubit state and subsequently measure the Q-function of the cavity mode (see Methods).  Figure 2B depicts the measured lattice site populations after each step of the walk. We observe the expected ballistic expansion of the coherent state in cavity phase space, consistent with theoretical predictions (population fidelities $>90$\%). 

As the walk unitary, $\hat{U}_W$, directly couples the particle's spin and position degrees of freedom, the resulting dynamics mimic those of spin-orbit interacting materials. More precisely, the unitary quantum walk protocol simulates continuous evolution under an effective spin-orbit Hamiltonian $\hat{H}_W$, which generates the same transformation as a single step of the walk when  $\hat{U}_W = e^{-i\hat{H}_W}$.  Since the unitary is translation invariant, the effective Hamiltonian exhibits Bloch bands of quasienergy $\pm\epsilon(k)$, where the quasimomentum, $k$, lies in the Brillouin zone; figures 3A and 3B show respectively the bandstructures underlying the walks $\hat{U}_0$ and $\hat{U}_1$.  The corresponding eigenstates consist of extended Bloch waves with spin polarization $\pm\vec{n}(k)$ \ \cite{Kitagawa2010}.  Depending on symmetry, the bandstructure of such spin-orbit-coupled Hamiltonians can feature quantized topological invariants.  In the case of the split-step quantum walk, $\Gamma = e^{-i \pi \vec{A} \cdot \vec{\sigma} /2 }$ plays the role of a so-called chiral symmetry\ \cite{Kitagawa2010, PhysRevB.88.121406}, with $\Gamma^\dagger \hat{U}_W  \Gamma = -\hat{U}_W$. This symmetry constrains the spin polarization vector $\vec{n}(k)$
 to lie on a great circle of the Bloch sphere, perpendicular to $\vec{A} = (\cos(\theta_1/2),0,\sin(\theta_1/2))$ (Fig. 3C and D).  Thus, the number of times $\vec{n}(k)$ wraps around the origin as $k$ varies through the Brillouin zone---known as the winding or Chern number $\mathcal{W}$---naturally defines the topological invariant\ \cite{Kitagawa2010} of the walk.  While the energy spectra of $\hat{U}_0$ and $\hat{U}_1$ are identical, they lie in topologically distinct phases, with $\hat{U}_0$ having zero winding number and $\hat{U}_1$ a winding number of unity.  Analogous to the number of twists in a closed ribbon, winding numbers are quantized and robust to local perturbations\ \cite{moore2010birth}.

\begin{figure}[h!]
\includegraphics[width=0.5\textwidth]{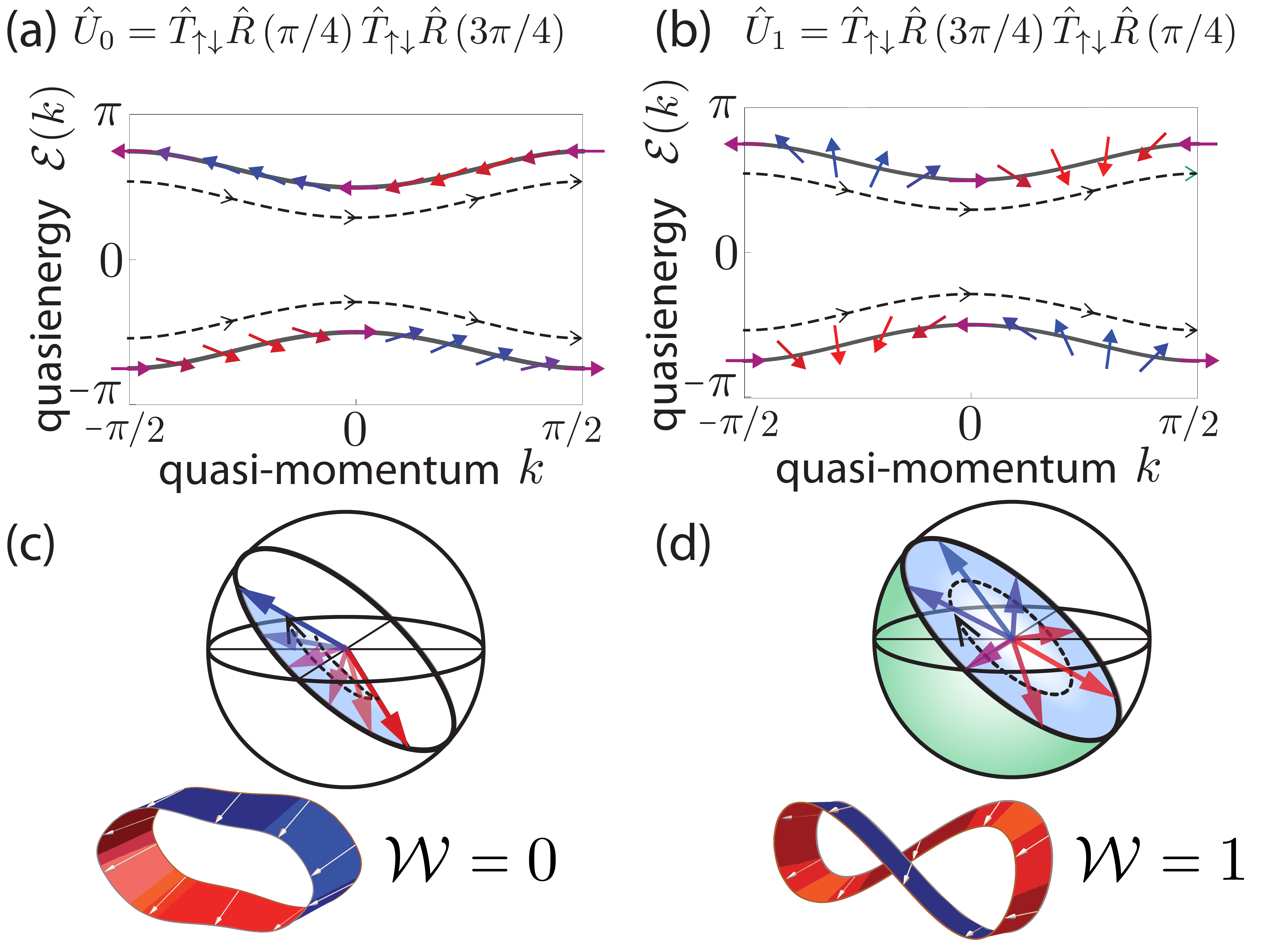}\caption{\textbf{Topological classes of  split-step quantum walks} Calculated band structures, quasienergy $\epsilon$ versus quasimomentum $k$, corresponding to the two walks we perform in the experiment, $\hat{U}_0 = \hat{T}_{\uparrow\downarrow}\hat{R}(\pi/4)\hat{T}_{\uparrow\downarrow}\hat{R}(3\pi/4)$ (\textbf{a}) and $\hat{U}_1 = \hat{T}_{\uparrow\downarrow}\hat{R}(3\pi/4)\hat{T}_{\uparrow\downarrow}\hat{R}(\pi/4)$ (\textbf{b}).     Though the energy bands of the two walks  are identical, they are topologically distinct, with the topology given by the winding of $\vec{n}(k)$ as $k$ varies through the Brillouin zone, shown in \textbf{c} and \textbf{d}. In \textbf{c}, the trivial case $\hat{U}_0$, $\vec{n}(k)$ does not complete a full revolution around the Bloch sphere, while in the topological case $\hat{U}_1$ \textbf{d}, it does perform a full revolution.  This also provides a direct connection to the Berry phase, as for a spin-1/2 system the Berry phase is simply half the subtended solid angle of the Bloch sphere path. A schematic representation of the variation of $\vec{n}(k)$ is shown by the ribbons below the Bloch spheres. The arrows on this strips point in the direction of $\vec{n}(k)$. Analogous to the number of twists in closed ribbons, winding numbers are quantized and robust to local perturbations.}
\end{figure}

The direct measurement of topological invariants in solid-state materials is an outstanding challenge\ \cite{Atala2013,aidelsburger2015measuring,roushan2014observation}, owing to the non-local nature of the order parameter. Our method makes use of a time-dependent modification of the quantum walk which, in the Hamiltonian picture, mimics an adiabatic translation of the underlying bandstructure across the Brillouin zone\ \cite{rama15,Banuls2006,Genske2013,matjeschk2012quantum,PhysRevLett.111.160601}.  The resulting dynamics effectively constitute digital Bloch oscillations, a phenomenon whereby a particle on a lattice subjected to a constant force undergoes oscillations~\cite{} due to the periodicity of the Brillouin zone.  In our system, these oscillations manifest as a refocusing of the quantum walker to its initial position, with a Berry phase---a signature of the bandstructure topology (see Fig. 3)---imprinted during the evolution.  In practice, this refocusing depends on choosing the number of steps in the walk such that the accrued dynamical phase---which has opposite signs in either band and thus impedes refocusing---effectively vanishes~\cite{supplement,rama15}. 
In the general setting, one can experimentally determine the condition for dynamical phase refocusing by performing spectroscopy while varying the number of steps.

In addition to the dynamical phase, upon traversing the Brillouin zone, the particle's spin winds around the Bloch sphere, encoding its path in the accumulated Berry phase\ \cite{Zak1989},
\begin{equation}
\phi_B=i\int_\mathrm{BZ}\langle k, \vec{n}(k)|\partial_k |k, \vec{n}(k)\rangle dk=\pi\times \mathcal{W}
\end{equation}
which thus becomes an observable manifestation of the winding number $\mathcal{W}$---the Hamiltonian's topological invariant. As one cannot directly observe the quantum mechanical phase of a wavefunction, measuring this Berry phase requires an interferometric approach. To this end, we perform the time-dependent walk with the cavity-qubit system initialized in a Schr\"odinger cat superposition of two coherent-state components: one component  undergoes the walk, while the other is unaffected by the  unitaries.  The Berry phase thus appears as the relative phase between the two components and is observable via direct Wigner tomography.  

The additional steps used in performing the time-dependent walks are shown in dashed boxes in Fig. 2A.  Beginning with either $\hat{U}_0$ or $\hat{U}_1$, we insert rotations by $\Delta k$ about  $\hat{\sigma}_z$  after each coin toss rotation $\hat{R}(\theta_1)$ and $\hat{R}(\theta_2)$.  In contrast to the original operations comprising $\hat{U}_0$ and $\hat{U}_1$, the rotation angle $\Delta k$ varies in time.  Since a $\hat{\sigma}_z$ rotation is equivalent to a translation of the underlying Hamiltonian in quasimomentum space\ \cite{rama15}, this time-varying rotation angle implements a digital Bloch oscillation. We choose $\Delta k$ to vary in steps of $\pi/10$ from $0$ to $\pi$, traversing the Brillouin zone exactly once.

Populations resulting from the time-dependent walks (with the system initialized in a single coherent state) are shown in Fig. 2C.  Unlike the ballistic dynamics resulting from the original walks, the Bloch oscillation (traversal of the Brillouin zone) causes the walker wavefunction to refocus\ \cite{Banuls2006,Genske2013,matjeschk2012quantum,PhysRevLett.111.160601} to both its initial position and spin state.  The intuition underlying this refocusing is that both the dynamical and Berry phase accumulated by each quasimomentum-component of the walker is identical upon full traversal\ \cite{rama15} (see Methods). In practice, we observe refocusing fidelities $>90 \%$, limited by incomplete adiabaticity and experimental imperfections.

\begin{figure}[h!]
\includegraphics[scale=0.2]{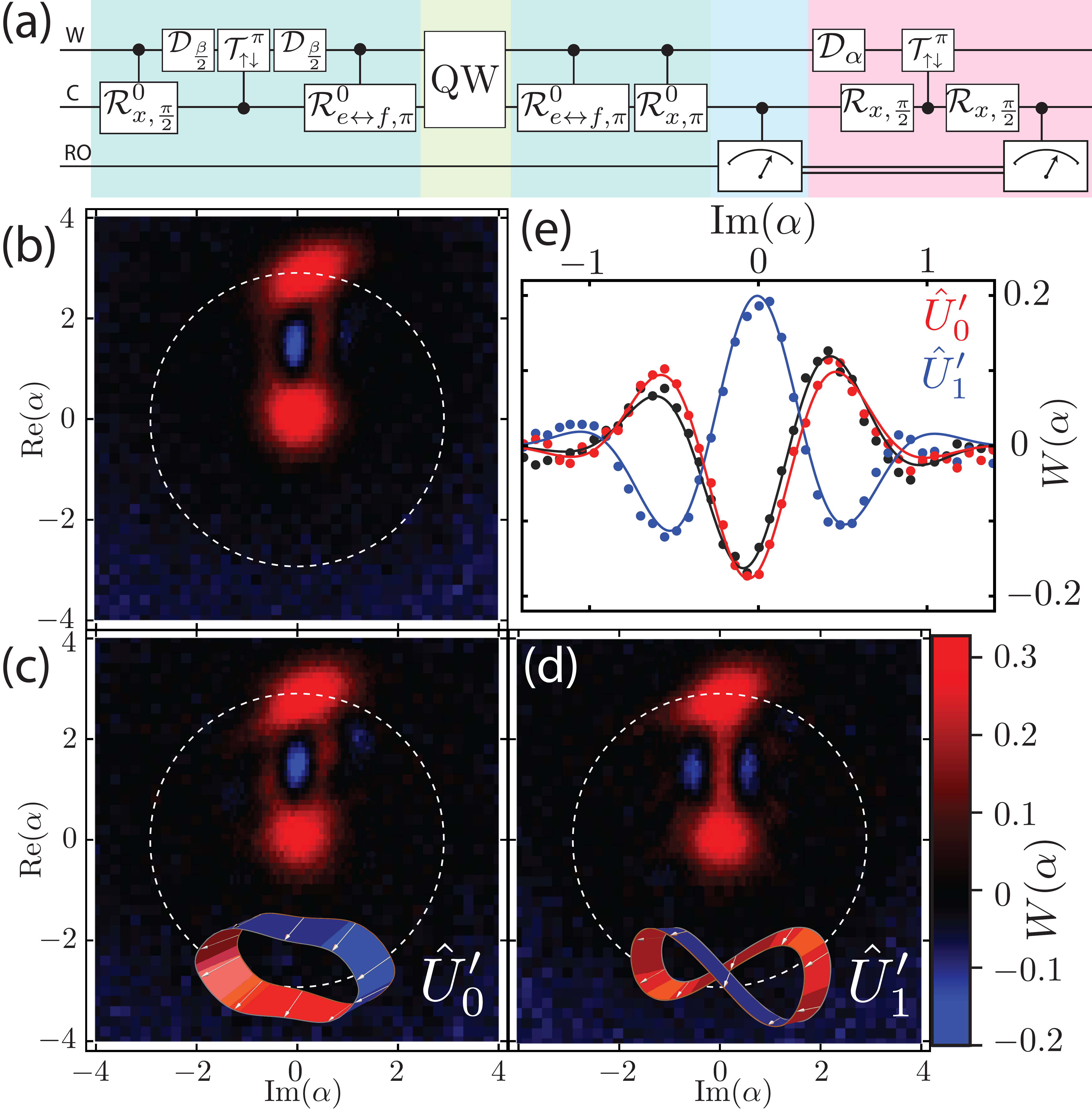}\caption{\textbf{Winding number measurement via direct Wigner tomography of refocused Schr\"odinger cat states.} \textbf{a.} Protocol for measuring topology via a time-dependent walk (Bloch oscillations). The Schr\"odinger cat state is first prepared (blue), after which the ten-step refocusing quantum walk is performed (green).  The qubit and cavity state are then disentangled, the qubit state is purified (blue), and direct Wigner tomography on the cavity state is performed (pink).  Wigner tomography of \textbf{b.} the cat undergoing no quantum walk, \textbf{c.} the cat after undergoing the trivial $\hat{U}_0$ walk, and \textbf{d.} after undergoing the topological $\hat{U}_1$ walk.  Fidelities of these resulting cat states compared to pure cat states are 0.68, 0.69, and 0.67 respectively.  \textbf{e.} A cut of the Wigner function, showing the fringes which encode the relative phase between the two cat components.  The Berry phase---captured by the phase difference between the topological and the trivial walks---is $\phi_B = 1.05\pi \pm 0.06\pi$ in experiment, consistent with the theoretical expectations of $\pi$. }
\end{figure}

Having verified the refocusing behavior of the time-dependent quantum walks, we initialize the cavity-qubit system in a Schr\"odinger cat state to measure the accumulated Berry phase (see Methods). One component of the cat is precisely the initial state of the previous walks, $|\beta,\uparrow\rangle$. The other component is $|0,f\rangle$, where the cavity in its ground (vacuum) state and the transmon in its second excited state\ \cite{Koch2007},  $|f\rangle$.  Shelving the vacuum component of the cat in the $|f\rangle$ state renders it immune to the coin toss rotations, as the $|f\rangle \leftrightarrow \left |\downarrow \right \rangle$ transition is far detuned ($225~\mathrm{MHz}$) from the $\left |\uparrow \right \rangle\leftrightarrow\left |\downarrow \right \rangle$ transition.  Thus, this component of the cat lies dormant during the walk, acting as a phase reference for the observation of the Berry phase. 
Our method of preparing the cat, a modification of the protocol introduced in ref.\ \cite{Vlastakis2013b}, is shown in Fig. 4A.  With the cat initialized, we perform the time-dependent walk over a full Bloch oscillation, applying the same set of pulses that resulted in the final frames of Fig. 2C.  After the walking component of the cat refocuses, we disentangle the qubit from the cavity with the operation $|0,f\rangle \rightarrow |0,\uparrow\rangle$.  This leaves the oscillator in the state
\begin{equation}
|\psi\rangle = |0\rangle - e^{i\phi_B}|\beta\rangle,
\end{equation}
where $\phi_B$ is the Berry phase.

While Q tomography lends itself well to measuring coherent state occupations, coherences between these states are largely invisible in this representation.  To measure the Berry phase, we therefore apply direct Wigner tomography to the final cavity state\ \cite{Raimond2006,Vlastakis2013b} (see Methods).  As figures 4 B, C, and D show, the Wigner functions of two-component cat states display interference fringes, whose phase directly encodes the relative phase between the dormant ($|0,f\rangle$) and walking ($|\beta,\uparrow\rangle$) components of the cat.
Figures 4 D and E display the measured Wigner functions for both split-step walks.  In the topologically trivial phase (Fig. 4D), the interference fringes do not acquire any phase shift after the walk, besides a small offset due to technical imperfections.  For the topologically nontrivial walk (Fig. 4E), however, the fringes visibly shift (Fig. 4C), corresponding to an acquired phase of $\phi_B = 1.05\pi \pm 0.06\pi$.  The topologies of the Hamiltonians which generate the walks are thereby clearly imprinted on the Wigner functions of the refocused states. A key feature of such topology is its robustness to all perturbations that do not close the spectral gap. To this end, we have performed an additional pair of quantum walks, $\hat{U}_0' = \hat{U}_W(0.64\pi,0.28\pi)$ and $\hat{U}_1' = \hat{U}_W(0.28\pi,0.64\pi)$ which are continuously deformable from the original walks. In this case, linecuts of the two Wigner functions yield an  extracted Berry phase difference of $\Delta\phi = 1.07\pi\pm0.09\pi$ ~\cite{supplement}. Thus, we have successfully observed, in a systematic fashion, both  phases in the canonical BDI topological insulator class\ \cite{schnyder2008classification,kitaev2009periodic}.

In conclusion, we have demonstrated a novel quantum-walk based simulator capable of emulating topological phases and directly measuring their topological invariants.  These invariants underlie phenomena such as topologically protected edge states\ \cite{kitagawa2012observation}, which have been previously observed with quantum walks.  In directly measuring the associated topological invariants, our work provides the missing piece of this bulk-edge correspondence for quantum walks.  A direct extension of our protocol is the realization of multi-dimensional quantum walks\ \cite{Kitagawa2010}, which has the potential to simulate novel topological insulators in two and three dimensions (e.g. Hopf insulator)\ \cite{Moore2008}. Looking forward, an outstanding challenge is the generalization of our protocol beyond single-particle physics, to measure topological invariants of interacting quantum many-body systems \cite{Schreiber2012}.  

\begin{acknowledgements}
The authors acknowledge discussions with David Toyli, Chris Macklin, Kevin Fischer, Mark Rudner, Eugene Demler, and Carlos Navarette-Benloch for motivating the use of a refocusing quantum walk.  VVR and LSM acknowledge funding via NSF graduate student fellowships. NYY acknowledges support from the Miller Institute for Basic Research in Science.   This research is  supported in part by the U.S. Army Research Office (ARO) under grant no. W911NF-15-1-0496 and by the AFOSR under grant no. FA9550-12-1-0378.
\end{acknowledgements}


\bibliography{biblio}
\bibliographystyle{ieeetr}

\pagebreak
\clearpage
\widetext
\begin{center}
\textbf{\large Supplemental Material : Observing Topological Invariants Using Quantum Walks in Superconducting Circuits}

\vspace{5mm}

{\large E. Flurin, V. V. Ramasesh, S. Hacohen-Gourgy, L. S. Martin, N. Y. Yao, I. Siddiqi}
\end{center}

\vspace{3mm}

\setcounter{equation}{0}
\setcounter{figure}{0}
\setcounter{table}{0}
\setcounter{page}{1}
\makeatletter
\renewcommand{\theequation}{S\arabic{equation}}
\renewcommand{\thefigure}{S\arabic{figure}}

This supplemental material expands on certain theoretical ideas from the main text, as well as giving a detailed account of the experimental setup.  Additionally, we present a measurement of the winding number taken at a pair of angles $\theta_1$ and $\theta_2$ incommensurate with those of the main text (namely $0.28\pi$ and $0.64\pi$.)  

\section{System description and Implementation of the Spin-Dependent Translation}

The Hamiltonian of the qubit-cavity system (not including the fundamental cavity mode, which is used for readout only) is the dispersive Jaynes-Cummings Hamiltonian\ \cite{Paik2011}: 
\begin{equation}
\hat{H}/\hbar=\omega_q \hat{\sigma}_z/2+\omega_c \hat{a}^\dagger\hat{a}+\chi  \hat{a}^\dagger\hat{a}\hat{\sigma}_z/2
\end{equation}
Effectively, this can be viewed as a cavity with a frequency which depends on the state of the qubit.  Therefore, the free evolution operator in the rotating frame of the cavity for a time $t$ corresponds to a spin-dependent phase shift given by
\begin{equation}
\hat{T}_{\uparrow\downarrow}=e^{i \phi \hat{a}^\dagger\hat{a}\hat{\sigma}_z}
\end{equation}
where the phase shift is $\phi=\chi t /2$.  Since the lattice is a set of coherent states in the phase space of the cavity, namely $\{\ket{x}=|\beta e^{i x\phi}\rangle,x\in\mathbb{Z}\}$ (where $\phi = 2\pi/10$), this unitary operation acts as a spin-dependent translation on the set of lattice states defined in the main text.  Explicitly,
\begin{equation}
\hat{T}_{\uparrow\downarrow}|\beta e^{i x\phi}\rangle|\uparrow \downarrow\rangle=|\beta e^{i (x\pm1)\phi}\rangle|\uparrow \downarrow\rangle
\end{equation}
Thus, the crucial spin-dependent translation gate is implemented by simply waiting for a time $t$ between successive rotations, where $t=2\phi/\chi=124\ \mathrm{ns}$.

\section{Cavity field representation}
Measurement of the system state after the walk, including both the lattice populations and their corresponding spin states, is performed via a sequence of two measurements: first, the qubit state is measured; second,  cavity state tomography is performed.  We perform two types of tomography of the cavity state to measure one of two phase space quasiprobability distributions:  either the Husimi Q-function or the Wigner function.  The Q function encodes the overlap of the resonator state $\ket{\psi}$ with a coherent state $\ket{\alpha}$: $Q_0(\alpha) = |\braket{\alpha |\psi}|^2/\pi$.  Since the lattice for our implementation of the walk is a set of coherent states, Q tomography is well suited to measure site populations.  However, the coherences between different states are exponentially suppressed in the Q function.  Wigner tomography, on the other hand, emphasizes coherences in the form of interference fringes.  Thus, for the Berry phase measurement discussed in the main text, we use Wigner tomography. Wigner tomography consists of measuring the parity $\Pi$ of the oscillator state after displacing it in phase space by an amount $\alpha$: $W(\alpha) = \braket{\psi| D_{\alpha}\Pi D_{-\alpha}|\psi}/\pi$.

\subsection{Husimi Q tomography}
To measure the Q function for a state $\ket{\psi}$, denoted $Q_{\ket{\psi}}(\alpha)$, we use the standard protocol~\cite{Kirchmair2013} with a minor modification.  $Q_{\ket{\psi}}(\alpha)$ is normally measured by displacing the oscillator by an amount $\alpha$, then measuring the probability for the cavity to contain $0$ photons.  This photon number measurement makes use of the qubit: a number-selective $\pi$ pulse is applied, mapping the zero-photon cavity population to the excited state of the qubit, whose state is then measured.  This process is done repeatedly, and the measurement statistics of the qubit give the zero-photon cavity population.  For this protocol to work, the qubit is assumed to start in the ground state, which in our case it does not due to the quantum walk.  In our experiment, the qubit-cavity system starts out in an entangled state $\ket{\psi} = c_\uparrow\ket{\psi_\uparrow,\uparrow} +c_\downarrow \ket{\psi_\downarrow,\downarrow}$.  So we first displace the state by $\ket{\alpha}$, and then measure the state of the qubit, picking out the displaced excited or ground state wavefunction.  Then we apply a selective $\pi$ pulse to the qubit and measure its state. 
Each frame of the Q functions shown in figure 3 of the main text consists of a 41-by-41 grid of displacements.  In total, 12 million measurements were acquired with a repetition rate of $500\ \mathrm{\mu s}$.  All displacements were performed at the cavity frequency corresponding to the qubit in the ground state, so to work in the desired frame (the rotating frame of the bare cavity frequency), the plots were rotated after acquisition.  Performing displacements at the bare cavity frequency was not practically feasible for technical reasons.

The distribution plotted in figure 3 of the main text corresponds to the superposition of the two weighted Q-functions, namely $|c_\uparrow|^2Q_{|\psi_\uparrow\rangle}(\alpha)$ and $|c_\downarrow|^2Q_{|\psi_\downarrow\rangle}(\alpha)$. The qubit populations $|c_\uparrow|^2$ and $|c_\downarrow|^2$ are extracted via the first QND readout. The detailed pulse sequence is shown in supplementary Fig.~\ref{fig:Sfig2}B.

\subsection{Wigner tomography}
To measure the Wigner function for a state $\ket{\psi}$, denoted $W_{\ket{\psi}}(\alpha)$, we again use the standard protocol~\cite{Vlastakis2013b} with a minor modification.  $W_{\ket{\psi}}(\alpha)$ is normally measured by displacing the oscillator by an amount $\alpha$, then measuring the parity of the resulting state.  This parity measurement is accomplished using the qubit, by applying two unconditional $\pi/2$ pulses separated in time by an amount $1/2\chi$.  Qubit measurement statistics give the state parity.  Our modification to this protocol, like in the Q-function measurement, puts the cavity displacement before the initial readout to project the qubit.    
Each frame of figure 4 of the main text, with the cat Wigner functions, consists of two sets of data, one coarse and one fine, meshed together. We zoomed in on the cat, taking 50 million measurements on a 41-by-41 grid of tomographic displacements corresponding to the $\alpha = [-1.25,1.25]\times[-0.75,3.75]$ region. The coarse set of data, superimposed, consists of another 41-by-41 grid in the $[-4,4]\times[-4,4]$ region with a total of 20 million measurements. We used a repetition rate of $500\ \mathrm{\mu s}$. The detailed pulse sequence is shown in Fig.~\ref{fig:Sfig2}(b).

\subsection{Population fidelity}
Populations $P(x,\uparrow \downarrow)$ on each lattice site $|x\rangle$ for the qubit state $|\uparrow \downarrow\rangle$ were extracted from the Q tomography using a Gaussian fit at the expected lattice position. The population fidelity to the theorical expectation $P_{\mathrm{th}}(x,\uparrow \downarrow)$ is then calculated using the following relation
\begin{equation}
F_{\mathrm{pop}}=\sum_x \sqrt{P(x,\uparrow)P_{\mathrm{th}}(x,\uparrow)+P(x,\downarrow)P_{\mathrm{th}}(x,\downarrow)}.
\end{equation}
This definition of population fidelity is similar to the standard definition of quantum fidelity, but it only considers the populations and disregards the coherences, which we do not direcly measure using Q tomography.  

\subsection{Wigner tomography fidelity}
The fidelities of the cat states to a pure target state $| \phi \rangle$ have been directly calculated from the Wigner tomography according to the following relation~\cite{Vlastakis2013b}
\begin{equation}
F=\langle \phi | \rho | \phi \rangle =\dfrac{1}{\pi}\int W_{\rm{target}}(\alpha) W_{\rm{cat}}(\alpha) \rm{d}^2\alpha
\end{equation}
where $W_{\rm{target}}(\alpha)$ is the Wigner function of the target state $|\psi \rangle$ and $W_{\rm{cat}}(\alpha)$ is the measured Wigner tomography of the cat states associated to the density matrix $\rho$.

\subsection{Errors associated with Wigner tomography measurement}
As Wigner tomography is a parity measurement of the displaced cavity state, the measurement of which is done via a qubit measurement, associated error bars can be calculated from a binomial distribution.  Here the probability of success is simply determined by the parity itself (the fraction of qubit measurements which collapsed to the excited state), calculated from the measured Wigner tomography value as $p=(1+\pi W(\alpha))/2$.  After $n$ tries, the  standard-error associated with binomial distribution is given by $\sigma=2/\pi \sqrt{p(1-p)/n}$.

We performed $5\times 10^7$ projective measurements per Wigner tomography on a 41-by-41 grid, resulting in $n=3\times 10^4$ projective measurements per point. Thus, the standard error is bounded by $1.4\times 10^{-3} <\sigma< 1.8\times 10^{-3}$. The radius of the points in the figure, $6\times10^{-3}$, is slightly larger the error bar associated with the noise due to projective measurements.  Thus, we have not shown any error bars in the figure.

\section{Bloch oscillation implementation}

Here, we expand on the discussion of Bloch oscillations in the main text.  Starting with the time-independent split step quantum walk $\hat{U}_W$ with unitary 
\begin{equation}
\hat{U}_W(\theta_1,\theta_2) = T_{\uparrow\downarrow} R(\theta_1) T_{\uparrow\downarrow} R(\theta_2),
\end{equation}
in terms of the quasimomentum operator $\hat{k} = \int_{BZ} k |k\rangle\langle k|$, this can be written
\begin{equation}
\hat{U}_W(\theta_1,\theta_2) = 
\exp{(i\hat{k}\sigma_z)}R(\theta_1)\exp{(i\hat{k}\sigma_z)}R(\theta_2).
\end{equation}
The underlying Hamiltonian stems $\hat{H}_W$ from this expression, by equating it to $\exp{(-i\hat{H}_W t/\hbar)}$.  Shifting the underlying Hamiltonian in quasimomentum space to implement the digitized Bloch oscillation can thus be accomplished by replacing $\hat{k}$ with $\hat{k}+\Delta k$ in the above expression: 
\begin{eqnarray}
\hat{U}_W &=& \exp{(i(\hat{k}+\Delta k)\sigma_z)}R(\theta_1)\exp{(i(\hat{k}+\Delta k)\sigma_z)}R(\theta_2) \\
&=& \exp{(i\hat{k}\sigma_z)}\exp{(i\Delta k\sigma_z)}R(\theta_1)\exp{(i\hat{k}\sigma_z)}\exp{(i\Delta k\sigma_z)}R(\theta_2).
\label{eq:BlochOsc}
\end{eqnarray}
To sweep out the entire Brillouin zone, we make the momentum shift in $\hat{U}_W$ vary at each step.  Choosing a value of $N$, the discretization of the Brillouin zone, we sweep the shift in units of $\Delta k = \pi/N$.  In our case, $N = 10$.

In practice, the $\sigma_z$ rotation is realized by composing a $\pi$-rotation about $\sigma_x$ with another $\pi$-rotation about $\vec{n}=\cos \frac{\Delta k}{2} \vec{x} -\sin\frac{\Delta k}{2} \vec{y}$ by virtue of the identity
\begin{equation}
e^{\frac{i}{2}\Delta k\sigma_z}=e^{\frac{i}{2}\pi(\cos \frac{\Delta k}{2}\sigma_x -\sin\frac{\Delta k}{2} \sigma_y)}e^{-\frac{i}{2}\pi\sigma_x}
\end{equation}
In practice, we contract two $\sigma_x$ rotations, the quantum walk rotation $R_x(\theta)$ with the first Bloch  rotation $R_x(-\pi)$. We can thus implement the quantum walk with Bloch oscillation by applying only two rotations in a row which reads
\begin{equation}
R_{n}(\pi)R_x(\theta-\pi)=R_{z}(\Delta k)R_x(\theta)
\end{equation}

In the ideal case, the value of $\Delta k$ within a single step would be constant, as shown in~\ref{eq:BlochOsc}.  However, it turns out to be more feasible experimentally to make $\Delta k$ vary smoothly over the entire protocol, so that instead of implementing at the $n$th step the unitary 
\begin{equation}
\hat{U}_W(n) = \exp{(i\hat{k}\sigma_z)}\exp{(in\Delta k\sigma_z)}R(\theta_1)\exp{(i\hat{k}\sigma_z)}\exp{(in\Delta k\sigma_z)}R(\theta_2),
\end{equation}
we actually implement
\begin{equation}
\hat{U}_W(n) = \exp{(i\hat{k}\sigma_z)}\exp{(in\Delta k\sigma_z)}R(\theta_1)\exp{(i\hat{k}\sigma_z)}\exp{(i(n+1/2)\Delta k\sigma_z)}R(\theta_2),
\end{equation}
Simulations show that for the value of $N=10$ and $N=12$ we use in the experiment, the behavior of this walk and the ideal are essentially identical.  

\section{Pulse sequence}

In this experiment, we use fixed frequency interactions and qubit/cavity drives in order to performs the requisite gates. We can perform both conditional and unconditional qubit/cavity operations by controlling the duration and shape of the driving fields used. The pulse sequences are presented in supplementary figures 2 and 3.

\subsection{Unconditional rotations}

Implementing the quantum walk consists of unconditional qubit rotations interspersed with spin-dependent translations corresponding to the free-evolution under the dispersive JC Hamiltonian for a time $\delta t= 2\pi / (5 \chi)=124\ \mathrm{ns}$. Qubit rotations have to be much shorter than this time to remain unconditional with respect to the cavity state.
In practice, we use $7.4\ \mathrm{ns}$ cosine-shaped pulses for unconditional qubit rotations. In order to prevent interaction and leakage to the $\ket{f}$ state, the second excited state of the transmon, we implement pulse-shaping techniques developed by\ \cite{Chen2015}, particularly using the Derivative Reduction for Adiabatic Gate (DRAG) pulse combined with a static detuning of $13.5\ \mathrm{MHz}$. In our experiment, these short pulses are generated using a single channel Arbitrary Wavefrom Generator (Tektronix 615) clocked at $2.7\ \mathrm{GS.s^{-1}}$ and modulated at $675\ \mathrm{MHz}$. Note that we use similar $\pi/2$-pulses to perform the parity measurement for the Wigner tomography.

\subsection{Conditional rotations}

Conditional rotations are used to prepare the cat state. A number-selective rotation is enabled by the dispersive Hamiltonian, particularly the fact that the qubit transition frequency depends on the photon occupation number of the cavity through $\omega_q^n=\omega_q^0-n\chi$.
By addressing individually one of these transitions, one can perform a photon-number-resolved qubit-rotation $R^n(\theta)=R(\theta)\otimes|n\rangle\langle n|+\mathbb{I}\otimes (\mathbb{I} -|n\rangle\langle n|)$. The selectivity of the rotation directly depends on the spectral selectivity of the pulse. For Gaussian pulses $\epsilon(t)=A e^{-t^2/(2\sigma)}$ one has to ensure that $\sigma \gg \chi^{-1}=100\ \mathrm{ns}$ to achieve a fully selective rotation. In our experiment, the Q-tomography is performed with a $\pi$-rotation highly selective with respect to the vacuum state $|0\rangle$: we use $\sigma=250\ \mathrm{ns}$ with a modulation frequency of $675\ \mathrm{MHz}$.

However, for the cat state preparation, the constraint is less strict since we want to perform a vacuum-state-selective rotation with respect to a coherent state separated by an amplitude of $\beta=2.78$, for which the first Fock state occupation remains small~\cite{Vlastakis2013b}.  Therefore, we perform selective pulses with a high fidelity using a Gaussian shape with $\sigma=63\ \mathrm{ns}$, and with a modulation frequency of $675\ \mathrm{MHz}$. Note that the selective $\pi$-rotation on the $\ket{e}\leftrightarrow \ket{f}$ transition is performed with the same pulse timing with a modulation frequency of $450\ \mathrm{MHz}$. These shorter pulses enabled us to dramatically increase the fidelity of the cat state preparation by mitigating errors originating from dephasing and the cavity's self-Kerr interaction~\cite{Kirchmair2013}.

\subsection{Cavity displacement}

Cavity displacements are unconditional with respect to the qubit state. We use Gaussian pulses with a width $\sigma=10\ \mathrm{ns}$. Cavity displacements are generated by a two-channel arbitrary waveform generator (Tektronix AWG520) clocked at $1\ \mathrm{GS.s^{-1}}$ and modulated at $125\ \mathrm{MHz}$.  In practice, we perform the cavity displacement for the Q and Wigner tomography before the first qubit projection. This enables us to avoid the deformation of the cavity state by the self-Kerr interaction during the readout time ($2.4\ \mathrm{\mu s}$) and further allows us to avoid interaction with the readout mode distorting the cavity state. Since the cavity lifetime ($40\ \mathrm{\mu s}$) is much larger than the readout time, the Q and Wigner tomography are not affected by this operation.

\subsection{Readout Pulse}

Three readouts are performed within a single pulse sequence: a first readout is used to herald the ground state of the qubit at the beginning of the experiment, a second one is performed for projecting the readout after the walk, and a last one is used for Q or Wigner tomography.
The readout pulses have been optimized for minimizing the readout time while achieving a high readout fidelity. The shape of the pulses can be decomposed into three components: a high amplitude gaussian rise to quickly populating the readout mode followed by a  medium amplitude plateau to hold the readout photon number during the actual measurement and finally a large Gaussian rise with opposite phase to quickly depopulate the readout mode. The relative amplitudes of the three components have been optimized such that the total readout time is $2.4\ \mathrm{\mu s}$ including the full depopulation time while achieving a readout fidelity of $F=96\ \%$. Note that the actual recording time, corresponding to the holding plateau, lasts $800\ \mathrm{ns}$. Readout pulses are generated by a two-channel AWG (Tektronix 520) clocked at $1\ \mathrm{GS.s^{-1}}$ without modulation.

\section{Device parameters}
\subsection{Transmon qubit}
The qubit consists of two aluminium paddles connected by a double-angle-evaporated aluminium Josephson junction deposited on double-side-polished sapphire. The resistance of the Josephson junction at room temperature is $7.1\ \mathrm{k\Omega}$.

\subsection{Superconducting cavity}

The qubit-cavity system is shown supplementary Fig. 5. The superconducting cavity is a $3"\times0.96"\times0.2"$ rectangular cavity made of high-purity aluminum (5N)\ \cite{Paik2011}. Three access ports are symmetrically positioned in the cavity. The readout port in the center strongly couples the $TE_{110}$ mode (readout) to a $50\ \Omega$ line through a non-magnetic SMA connector inserted in the cavity.  The pulsing port on one side of the cavity is very weakly coupled to the $TE_{210}$ mode in order to preserve its high quality factor while being able to address it. The port on the other side is not used but it is crucial to keep the overall symmetry of the cavity. Two slots are positioned symmetrically on each side of the cavity. One slot hosts the qubit chip while the other one hosts a blank sapphire chip with the same dimensions. The blank chip enables us to enforce the symmetry of the cavity modes and therefore to preserve the high-quality factor ($2\times10^6$) of the TE210 mode despite the strongly coupled readout pin sitting in the center. 

\subsection{Parametric amplifier}
The lumped-element Josephson parametric amplifier (LJPA)\ \cite{Hatridge2011} used to increase readout fidelity consists of a two-junction SQUID, formed from $2\ \mathrm{\mu A}$ Josephson junctions shunted by $3 \ \mathrm{pF}$ of capacitance, and is flux biased to provide $20\ \mathrm{dB}$ of gain at the cavity resonance frequency. The LJPA is pumped by two sidebands equally spaced $300\ \mathrm{MHz}$ above and below the cavity resonance.

\begin{center}
    \begin{tabular}{| l | l | l  |}
    \multicolumn{2}{c}{qubit}       \\ \hline
    $\omega_q/2\pi$& $5.186\ \mathrm{GHz}$ \\ \hline
    $T_1$ & $30\ \mathrm{\mu s}$  \\ \hline
   $T_{2}^*$ & $5.3\ \mathrm{\mu s}$   \\ \hline
    $T_{2}$ & $9\ \mathrm{\mu s}$ \\\hline
        $\alpha$& $225\ \mathrm{MHz}$ \\ \hline
    $\chi_{\rm{RO}}/2\pi$& $1.1\ \mathrm{MHz}$ \\ \hline
    $\chi_{\rm{m}}/2\pi$& $1.6125\ \mathrm{MHz}$ \\ \hline
    \multicolumn{2}{c}{ $TE_{210}$ mode }       \\ \hline
    $\omega_m/2\pi$& $7.414\ \mathrm{GHz}$ \\ \hline
    $T_1$ & $40\ \mathrm{\mu s}$  \\ \hline
    $Q$ & $2\times 10^6$   \\ \hline
    $\text{self-Kerr}\ K/2\pi$ & $\sim 3\ \mathrm{kHz}$ \\\hline
    $\chi_{\rm{m}}/2\pi$& $1.6125\ \mathrm{MHz}$ \\ \hline
      \multicolumn{2}{c}{ $TE_{210}$ mode (readout) }         \\ \hline
    $\omega_m/2\pi$& $6.767\ \mathrm{GHz}$ \\ \hline
    $\kappa_{ext}$ & $(270\ \mathrm{ns})^{-1}$  \\ \hline
    $Q$ & $10^4$   \\ \hline
    $\chi_{\rm{RO}}/2\pi$& $1.1\ \mathrm{MHz}$ \\ \hline
    \end{tabular}
\end{center}

\section{Topological Features of Quantum Walks}

We expand on the topological features of quantum walks discussed in the main text, directing the reader to refs.\ \cite{Kitagawa2010,kitagawa2012observation} for the original exposition.  Our definition of the split-step quantum walk discussed in the main text differs slightly from that of refs.\ \cite{Kitagawa2010,kitagawa2012observation}: we define the split-step quantum walk to consist of repeated applications of the operator
\begin{equation}
U\left(\theta_1, \theta_2\right) = T_{\uparrow \downarrow}R_x\left(\theta_1\right)T_{\uparrow \downarrow}R_x\left(\theta_2\right).
\end{equation}
where $R_x\left(\theta\right) = \cos{\left(\theta/2\right)}\hat{I} - i\sin{\left(\theta/2\right)}\sigma_x$ is a rotation operator acting only on the spin degrees of freedom, and 
\begin{equation}
 T_{\uparrow \downarrow} = \sum_x \left[\ket{x+1}\bra{x} \otimes \ket{\uparrow}\bra{\uparrow} + \ket{x-1}\bra{x} \otimes \ket{\downarrow}\bra{\downarrow} \right]
\label{eq:TOp}
\end{equation}
is the spin-dependent translation operator.  A Hamiltonian $H_W$ exists such that 
\begin{equation}
e^{-iH_\mathrm{W}\Delta t/\hbar} = U\left(\theta_1,\theta_2\right),
\label{eq:equivalence}
\end{equation} 
where $\Delta t$ is the interval between successive applications of $U\left(\theta_1, \theta_2\right)$.  The dynamics of a system evolving continuously under the steady-state Hamiltonian $H_\mathrm{W}$ are the same as the quantum walk dynamics if the state of the system is probed only at integer multiples of $\Delta t$. We take $\Delta t/\hbar=1$.  

The unitary operator performing the walk commutes with translations of the system by an integer number of lattice sites, so the stationary states of the walk are products of the spin-1/2 wave function and plane wave states $\ket{k}$ given by 
\begin{equation}
\ket{k} = \sum_x e^{ikx}\ket{x}.
\label{eq:planeWave}
\end{equation}  

Because the quantum walk protocol is invariant under discrete lattice translations, the Hamiltonian must be block diagonal in the quasimomentum basis; thus the most general form it can have is
\begin{equation}
 H(\hat{k})=\epsilon(\hat{k})\vec{n}(\hat{k}).\vec{\hat{\sigma}}+\gamma(\hat{k})I
 \label{eq:QWalkH1}
 \end{equation}
In the quantum walk protocols we consider, the term proportional to the identity (a quasimomentum-dependent band offset) turns out to zero, giving the form 
\begin{equation}
H(\hat{k})=\epsilon(\hat{k})\vec{n}(\hat{k}).\vec{\hat{\sigma}}.
\label{eq:QWalkH2} 
\end{equation}
Substituting this into eq.\ \ref{eq:equivalence} gives the expressions for $\epsilon(k)$ and $\vec{n}$:
\begin{eqnarray}
\cos{\epsilon(k)} &=&\cos{2k}\cos{(\theta_2/2)}\cos{(\theta_1/2)}+\sin{(\theta_2/2)}\sin{(\theta_1/2)}\nonumber \\
n_x(k) &=& \dfrac{\cos{2k}\cos{(\theta_2/2)}\sin{(\theta_1/2)}+\sin{(\theta_2/2)}\cos{(\theta_1/2)}}{\sin{\epsilon(k)}}\nonumber \\
n_y(k) &=& -\dfrac{\sin{2k}\cos{(\theta_2/2)}\sin{(\theta_1/2)}}{\sin{ \epsilon(k)}}\nonumber \\
n_z(k) &=& -\dfrac{\sin{2k}\cos{(\theta_2/2)}\sin{(\theta_1/2)}}{\sin{\epsilon (k)}}
\end{eqnarray}

For a particular walk (i.e. a particular value of $\theta_1$ and $\theta_2$), there exists a vector $\vec{A}$ such that $\hat{n}(k)$ is perpendicular to $\vec{A}$ for all $k$:
\begin{equation}
\vec{A} = -\cos{(\theta_1/2)}\vec{y} + \sin{(\theta_1/2)}\vec{z}
\end{equation}
This constraint forces $\vec{n}(k)$, a unit vector by definition, to lie along the great circle of the Bloch sphere perpendicular to $\vec{A}$.  $H_\mathrm{W}$ can thus be characterized by the number of times $\hat{n}(k)$ winds around the vector $\vec{A}$ as $k$ traverses the Brillouin zone.  

\subsection{Dynamical Phase in Quantum Walks}

As stated in the main text, the digital Bloch oscillation imprints a Berry (geometric) phase on the wavefunction.  However, in general, the Bloch oscillation also imprints a dynamical phase on the walker.  The existence of this dynamical phase $\phi_d$ prevents the state from refocusing unless $\phi_d$ is a multiple of $2\pi$, since unlike the Berry phase, the dynamical phase has opposite signs for states in different bands (that is, states in the upper band pick up a dynamical phase $+\phi_d$, while states in the lower band pick up a dynamical phase $-\phi_d$).  In continuous-time systems with energy bands given by $\epsilon(k)$, the dynamical phase $\phi_d$ is given by ($\hbar = 1$)
\begin{equation}
\phi_d = \int_{t_i}^{t_f}\epsilon(k(t))~dt.
\end{equation}
Assuming a linear traversal of the Brillouin zone with velocity $v = dk/dt$, the dynamical phase is simply related to the integral under the energy bands:
\begin{equation}
\phi_d = \frac{1}{v}\int_{\mathrm{BZ}} \epsilon(k)~dk 
\end{equation}
In the digital analogue of Bloch oscillations performed in the experiment, this formula takes the form 
\begin{equation}
\phi_d = \frac{N}{\pi}\int_{\mathrm{BZ}} \epsilon(k)~dk, 
\end{equation}
where $N$ is the number of steps in which the Brillouin zone traversal is discretized.  By choosing a suitable value of $N$, we can make the acquired dynamical phase arbitrarily close to a multiple of $2\pi$, so that it can be ignored.  For the bandstructure of our walks, with coin toss angles $\pi/4$ and $3\pi/4$, choosing $N = 10$ gives a dynamical phase which is close to a multiple of $2\pi$ and thus allows the state to refocus.  For more details, see\cite{rama15}.

\subsection{View of Evolution in Quasimomentum Space}
As the quantum walk is initialized with the walker localized on a single lattice site, the initial wavefunction is not an eigenstate of the effective Hamiltonian but rather a superposition of all the quasimomentum states allowed by the periodic boundary conditions of our system and their corresponding spin eigenstates.  That is, for those quasimomenta $k$ satisfying $e^{ikN}=1$, with $N$ the number of lattice sites in the circle, the initial wavefunction can be written 
\begin{equation}
|\Psi\rangle=\sum_k \alpha_k |k,\hat{n}(k)\rangle + \beta_k |k,-\hat{n}(k)\rangle,
\end{equation}
where $|\hat{n}(k)\rangle$ and $|-\hat{n}(k)\rangle$ are the spin eigenstates corresponding to the quasimomentum $k$, i.e. the spin eigenstates of the Hamiltonian $H = \hat{n}(k)\cdot\vec{\sigma}$.

Upon undergoing the refocusing quantum walk, the states in the upper and lower bands evolve as follows: 
\begin{eqnarray}
|k,\hat{n}(k)\rangle &\rightarrow& e^{+i\phi_d+i\phi_B^+} |k,\hat{n}(k)\rangle \\
|k,-\hat{n}(k)\rangle &\rightarrow&
e^{-i\phi_d+i\phi_B^-} |k,-\hat{n}(k)\rangle \end{eqnarray}
Here $\phi_d$ is the dynamical phase, which for the number of steps we have chosen, is a multiple of $2\pi$ and can be neglected; while $\phi_B^+$ and $\phi_B^-$ are the Berry phases corresponding to the upper and lower bands, given by (compare Eq. (2) in the main text)
\begin{eqnarray}
\phi_B^{\pm}=i\int_\mathrm{BZ}\langle k, \pm\vec{n}(k)|\partial_k |k, \pm\vec{n}(k)\rangle dk \\
\end{eqnarray}
As for general spin-1/2 systems, the Berry phase is equal to half the subtended solid angle of the path $\vec{n}(k)$ as $k$ traverses the Brillouin zone.  In our case, $\vec{n}(k)$ either winds around the Bloch sphere once or zero times; in either case, the winding of $\vec{n}(k)$ and $-\vec{n}(k)$ is the same, and thus the accumulated Berry phase is the same for both bands.  

\section{Additional Winding Number Measurement}

Here we present an additional measurement to the one described in the main text, in which we performed both time-independent and time-dependent split-step quantum walks with angles $\theta_1= 0.28\pi$ and $\theta_2= 0.64\pi$.  These results of these operations are shown in supplementary figure 1.  

Specifically, the unitary operations, $\hat{U}_0'$ and $\hat{U}_1'$,  for each step of these split-step walks are given by 
\begin{eqnarray}
\hat{U}_0' &=& \hat{T}_{\uparrow\downarrow} \hat{R}(0.28\pi) \hat{T}_{\uparrow\downarrow} \hat{R}(0.64\pi) \\
\hat{U}_1' &=& 
\hat{T}_{\uparrow\downarrow} \hat{R}(0.64\pi) \hat{T}_{\uparrow\downarrow} \hat{R}(0.28\pi)
\end{eqnarray}

In the measurement presented in the main text, we used a lattice with 10 sites; here, we use a lattice with 12 sites.  In figure 1A, we plot the free evolution under this quantum walk, with the state of the walker evolving for 12 steps.  As with the measurement in the main text, this serves primarily to benchmark our experimental platform, demonstrating its capability to perform the quantum walk algorithm.  The population fidelities of the final walker state to theoretical predictions are, similarly to the main text, quite high, above 93\% in all cases, and with an average of 0.97\% for both $\hat{U}_0'$ and $\hat{U}_1'$.  

Our time-dependent modification of the walk protocol which transports the walker around the Brillouin zone is, as in the main text, accomplished by inserting effective $\sigma_z$ rotations before each spin-dependent translation step.  Here, we use a time-dependent quantum walk with 12 steps.  In this case, the fidelity of refocusing is, respectively 82\% for $\hat{U}_0'$ and 80\% for $\hat{U}_1'$.

Using the Schr\"odinger cat, we again extract a measurement of the accumulated Berry phase, and hence the winding number, from a linecut of the data.  The corresponding plots are shown in panels C and D of supplementary figure 1, with the linecut shown in panel E.  As the panels show, there is, as in the main text, a striking difference in the Wigner functions between the topological and trivial walks.  We extract a Berry phase of $1.07\pi\pm 0.09\pi$.

\newpage

\begin{figure*}[h!]
\includegraphics[scale=0.15]{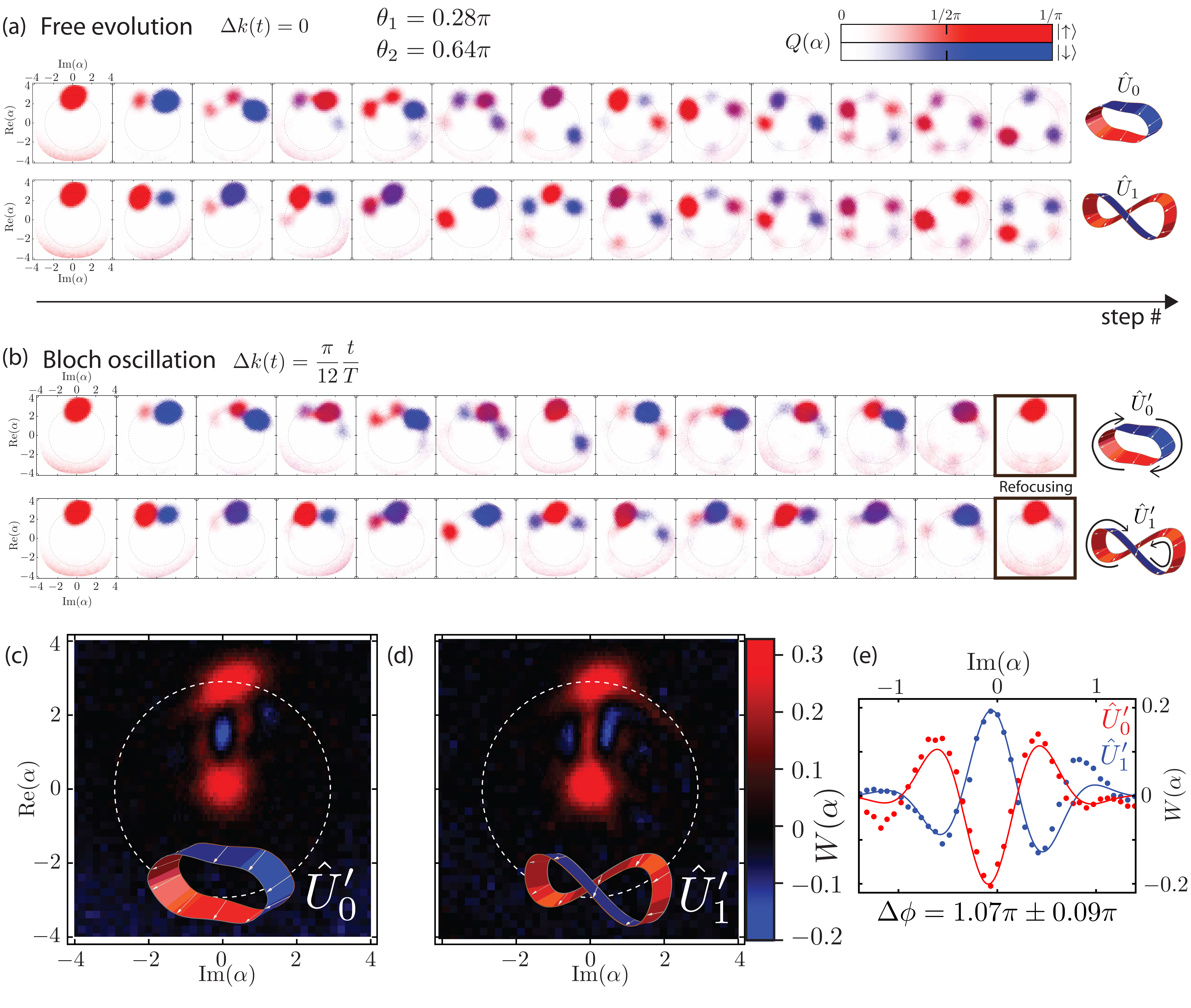}\caption{ \textbf{Quantum walk with angles $\theta_1 = 0.28\pi$ and $\theta_2 = 0.64\pi$} \textbf{a.}  Lattice site populations (cavity Q functions) after each step of the time-independent versions of $\hat{U}_0'$ (top strip) and $\hat{U}_1'$ (bottom strip).  Spin-up (red) and spin-down (blue) Q functions are superimposed, as with the walk performed in the main text.  Average fidelity of the populations compared to theoretical predictions is 0.97 for both $\hat{U}_0$ and $\hat{U}_1$. 
\textbf{b.} Cavity Q functions after each step of the refocusing quantum walk with Bloch oscillations.  The state refocuses after ten steps, as shown in the final frame for both $\hat{U}_0'$ and $\hat{U}_1'$.  Refocusing fidelities (to the initial state) for $\hat{U}_0'$ and $\hat{U}_1'$ are 0.82 and 0.80, respectively.
Wigner tomography of \textbf{c.} the initial cat, \textbf{d.} the cat after undergoing the trivial $\hat{U}_0'$ walk, and \textbf{d.} after undergoing the topological $\hat{U}_1'$ walk. \textbf{e.} Linecuts of the two Wigner functions, showing the extraction of the Berry phase difference, in this case $\Delta\phi = 1.07\pi\pm0.09\pi$. \label{fig:Sfig1}}
\end{figure*}

\begin{figure*}[h!]
\includegraphics[scale=0.2]{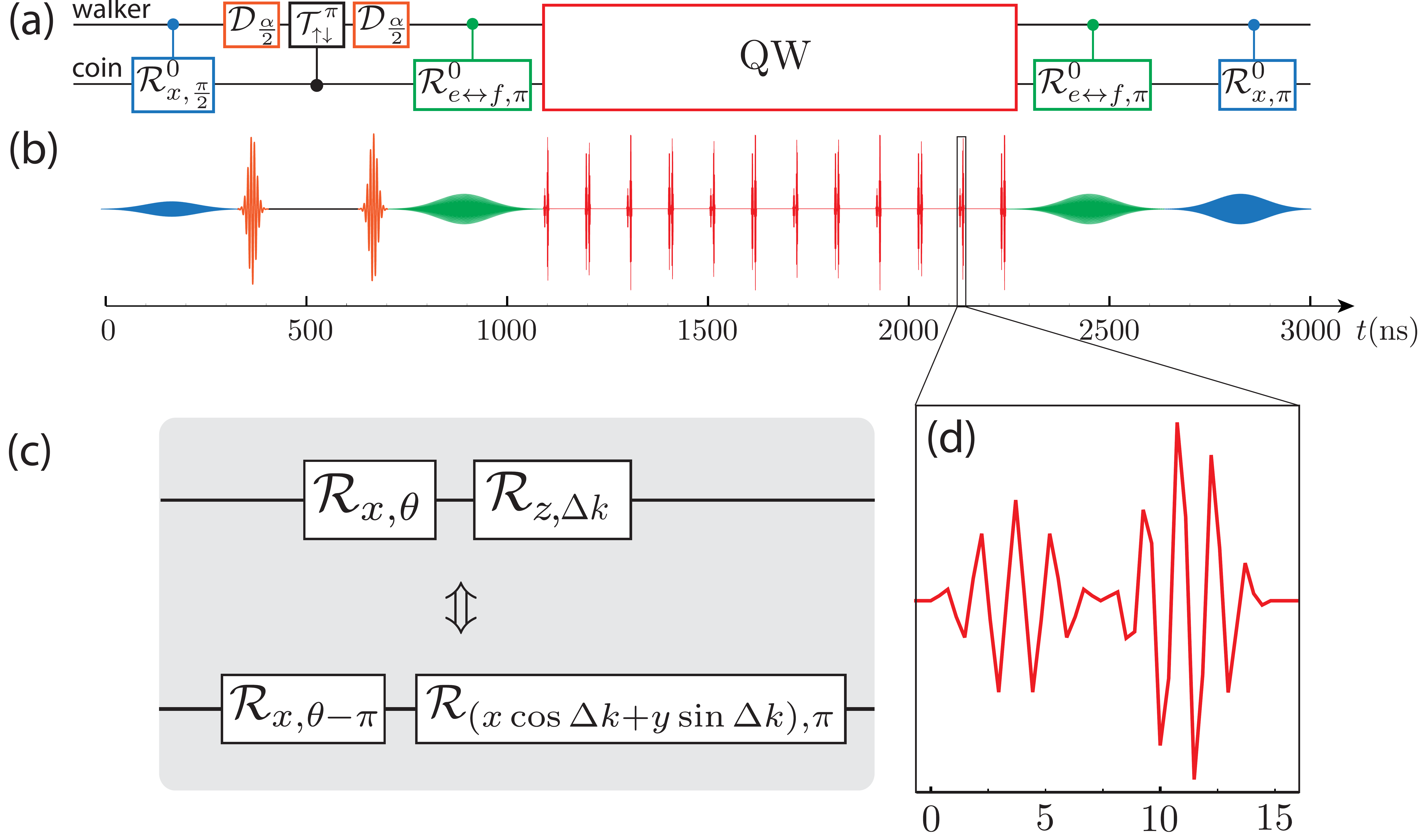}\caption{\textbf{Pulse shaping for quantum walk} \textbf{a.} Sequence of pulses which implement the quantum walk, including initialization of the Schroedinger cat state. \textbf{b.} Pulse shaping and timing corresponding to the sequence. \textbf{c.} Equivalence between two sets of composed rotations, with the bottom one implemented in the experiment.\textbf{d.} An expanded view showing the qubit rotation pulse shapes used for the time-dependent quantum walk.\label{fig:Sfig1}}
\end{figure*}

\begin{figure*}[h!]
\includegraphics[scale=0.2]{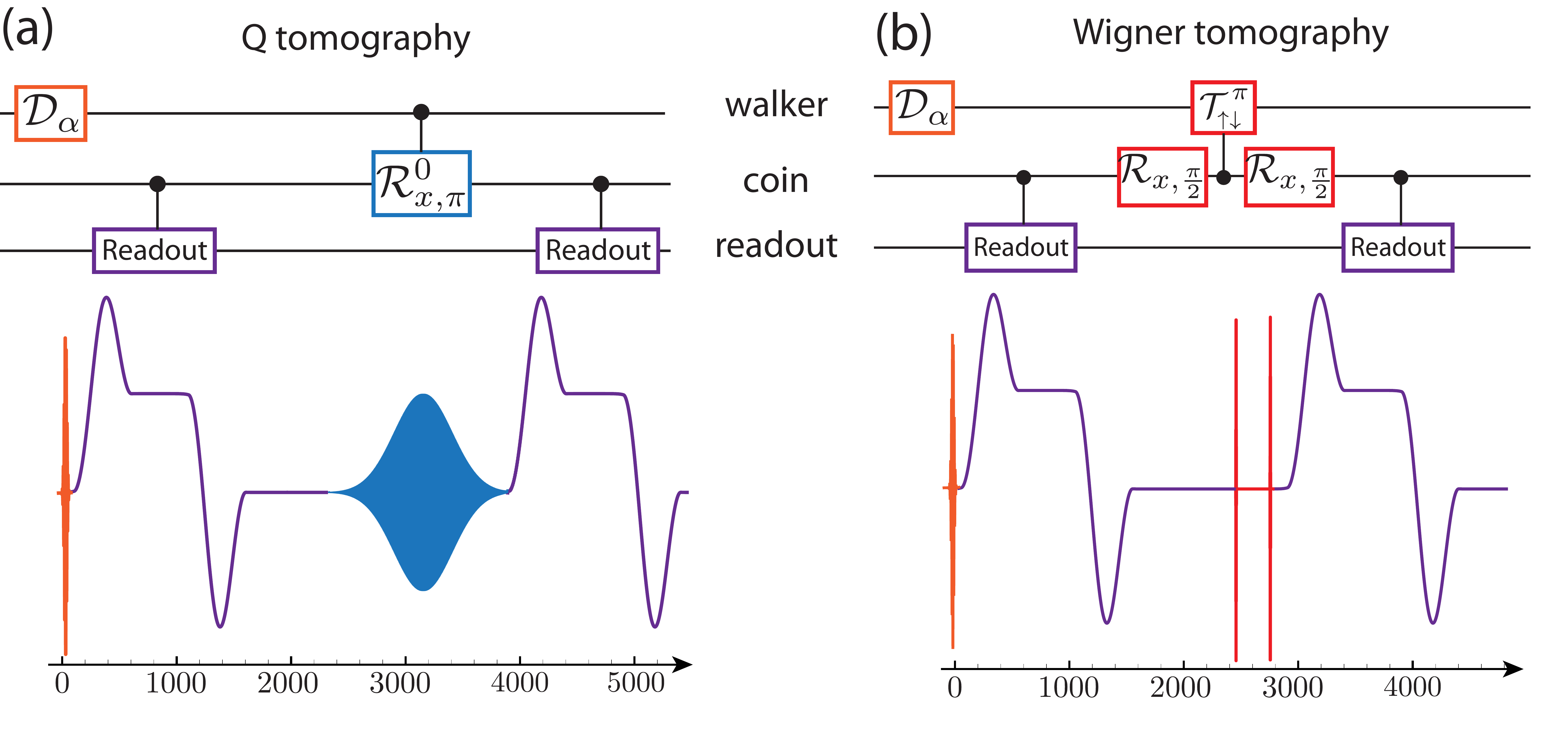}\caption{\textbf{Pulse shaping for Q and Wigner Tomography} \textbf{a.}  Gate sequence implemented to perform the Q tomography and the corresponding pulse sequence. \textbf{b.}Gate sequence implemented to perform the Wigner tomography and the corresponding pulse sequence.  \label{fig:Sfig2}}
\end{figure*}

\begin{figure*}[h!]
\includegraphics[scale=0.2]{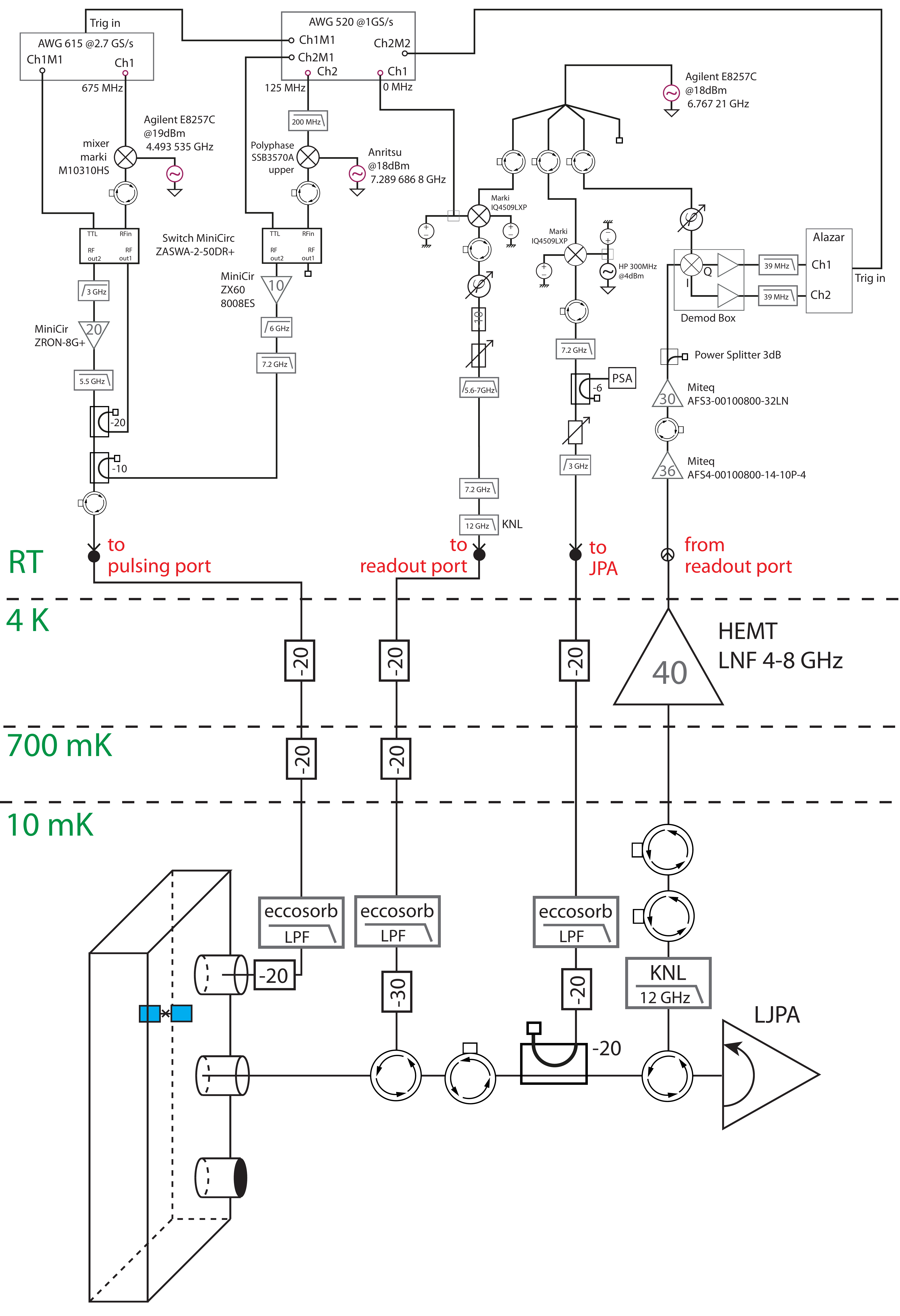}\caption{\textbf{Detailed block diagram of the measurement setup} Our measurement is performed in a dilution refrigerator at a base temperature of 10 mK.  We use two input lines.  One, directed into the weakly coupled pulsing port, serves three purposes: it has a high-amplitude channel for unconditional qubit pulses, a low-amplitude channel for conditional qubit pulses, and a channel for cavity displacement pulses. The other line is directed toward the readout port and is used to perform a homodyne reflection measurement of the $TE_{110}$ mode of the cavity. The readout signal is sent through a chain of low-noise amplifiers before down-conversion and digitization, allowing for qubit state measurement. A third line is used to pump the parametric amplifier with detuned sidebands.
\label{fig:Sfig3}}
\end{figure*}

\begin{figure*}[h!]
\includegraphics[scale=0.4]{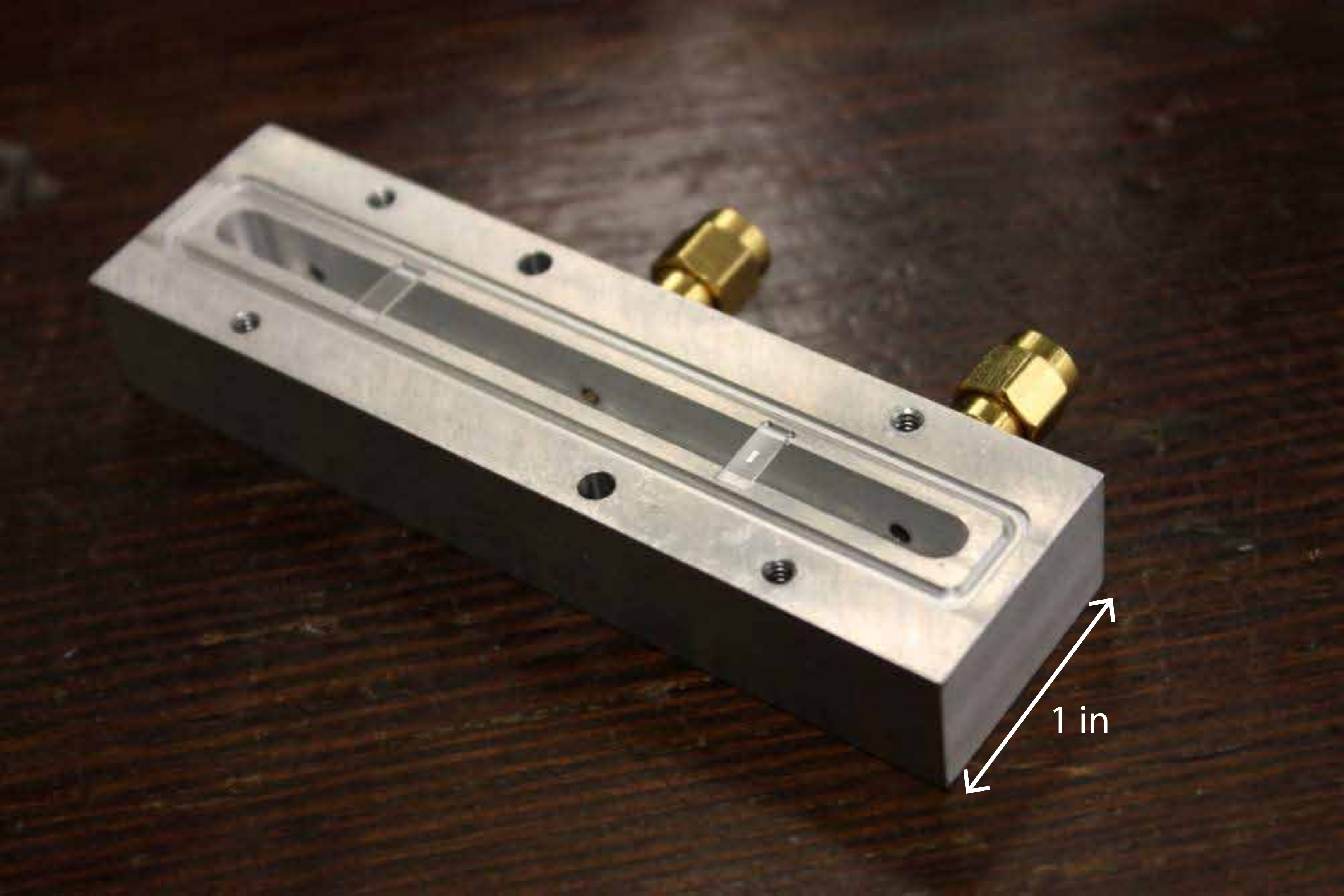}\caption{\textbf{Picture of the cavity embedding the superconducting transmon}
\label{fig:cavityPic}}
\end{figure*}

\end{document}